\documentclass[a4paper,10pt]{article}

\usepackage{amssymb}
\usepackage{amsmath}
\usepackage{graphicx}
\usepackage{hyperref}
\usepackage{amsthm}

\begin{document}

\bibliographystyle{alpha}

\title{$k$-Hyperarc Consistency for Soft Constraints \\over Divisible Residuated Lattices}

\author{Simone Bova\\
Department of Mathematics and Computer Science\\
University of Siena, Italy\\
\texttt{bova@unisi.it}}

\newtheorem{definition}{Definition}[section]
\newtheorem{theorem}{Theorem}[section]
\newtheorem{proposition}{Proposition}[section]
\newtheorem{example}{Example}[section]
\newtheorem{notation}{Notation}[section]
\newtheorem{lemma}{Lemma}[section]
\newtheorem{corollary}{Corollary}[section]
\newtheorem{fact}{Fact}[section]
\newtheorem{conjecture}{Conjecture}[section]
\newtheorem{claim}{Claim}[section]
\newtheorem{algorithm}{Algorithm}[section]

\maketitle

\begin{abstract}
We investigate the applicability of divisible residuated lattices (DRLs) as a general evaluation framework for soft constraint satisfaction problems (soft CSPs). DRLs are in fact natural candidates for this role, since they form the algebraic semantics of a large family of substructural and fuzzy logics \cite{GJKO07,H98}. 

We present the following results. $(i)$ We show that DRLs subsume important valuation structures for soft constraints, such as commutative idempotent semirings \cite{BMR97} and fair valuation structures \cite{CS04}, in the sense that the last two are members of certain subvarieties of DRLs (namely, Heyting algebras and $BL$-algebras respectively). $(ii)$ In the spirit of \cite{LS04,BG06}, we describe a polynomial-time algorithm that enforces $k$-hyperarc consistency on soft CSPs evaluated over DRLs. Observed that, in general, DRLs are neither idempotent nor totally ordered, 
this algorithm amounts to a generalization of the available algorithms that enforce $k$-hyperarc consistency. 
\end{abstract}

\section{Introduction}

A constraint satisfaction problem (CSP) is the problem of deciding, given a collection of constraints on variables, whether or not there is an assignment to the variables satisfying all the constraints. 
In the \emph{crisp} setting \cite{M74}, any assignment satisfying all the constraints provides a solution, 
and all the solutions are equally suitable. 
In the \emph{soft} setting \cite{BMRSVF99}, more generally, each constraint maps the assignments to a 
\emph{valuation structure}, which is a bounded poset equipped with a suitable \emph{combination} operator; 
the task is to find an assignment such that the combination of its images under all the constraints 
is maximal in the order of the valuation structure (formal definitions are given in Section~\ref{sect:softCSPs}). 
In general, the soft CSP is NP-complete, so that research efforts are aimed to characterize tractable cases \cite{CCJK06,DBLP:conf/cp/CohenCJ06}, and to describe polynomial-time \emph{enforcing} (or \emph{filtering}) 
algorithms. A tipical enforcing algorithm takes as input a soft CSP, and enforces a \emph{local consistency} property over the input problem, producing two possible outcomes: either the input problem is found locally inconsistent, implying its global inconsistency; or else, the input problem is transformed into an \emph{equivalent} problem, maybe inconsistent but \emph{easier}, that is, with a smaller solution space. 
Despite their incompleteness as inconsistency tests, enforcing algorithms are useful as subprocedures in the exhaustive search for an optimal solution, for instance in \emph{branch and bound} search \cite{D03}.

The generalization of local consistency notions and techniques from the crisp to the soft setting plays a central role in the algorithmic investigation of soft CSPs, and any class of structures that allows for an easy migration of local consistency techniques in the soft setting deserves consideration \cite{BMR97,LS04,CS04}. Not surprisingly, the \emph{weaker} the properties of the valuation structure are, the harder it is to migrate a local consistency technique from the crisp to the soft setting: indeed, loosely speaking, a crisp CSP is equivalent to a soft CSP over a valuation structure with very \emph{strong} properties, basically the algebra $(\{0,1\},\leq,\odot,\bot,\top)$, where $\bot=0 \leq 1=\top$ and $x \odot y = 1$ if and only if $x=y=1$. As a minimal requirement, a valuation structure for a soft CSP has to be a bounded poset, with top element $\top$ and bottom element $\bot$, equipped with a commutative, associative operation $x \odot y$ which is monotone over the order ($x \leq y$ implies $x \odot z \leq y \odot z$), has $\top$ as identity ($x \odot \top=x$) and $\bot$ as annihilator ($x \odot \bot=\bot$). Intuitively, an assignment mapped to $\top$ by a constraint is entirely satisfactory, and an assignment mapped to $\bot$ is entirely unsatisfactory; if two assignments are mapped to $x$ and $y$, in the case $x \leq y$, the latter is preferred to the former, 
whether in the case $x \parallel y$, they are incomparable; the operator $\odot$ combines constraints in such a way that adding constraints shrinks the solution space (as boundary cases, $\top$ does not shrink the solution space, 
and $\bot$ empties the solution space). In this setting, two options arise. The first is whether or not to allow incomparability, formally, whether or not to admit \emph{non-totally} ordered valuation structures; the second is whether or not to keep into account repetitions, formally, whether or not to allow for valuation structures with \emph{nonidempotent} combination operators (in the idempotent case $x \odot x=x$, so that repetitions do not matter).  The aforementioned algebra $(\{0,1\},\leq,\odot,\bot,\top)$ is strong in the sense that it is totally ordered and idempotent. 

In this paper, we propose \emph{(commutative bounded) divisible residuated lattices} (in short, \emph{DRLs}) as a unifying evaluation framework for soft constraints, and we provide two evidences supporting this proposal. The first is that DRLs (in general lattice ordered and nonidempotent) subsume important valuation structures where local consistency techniques succeeded, namely commutative idempotent semirings (lattice ordered and idempotent, \cite{BMR97}) and fair valuation structures (totally ordered and nonidempotent, \cite{CS04}). The second is that DRLs readily host a polynomial-time algorithm that enforces a useful local consistency property, called $k$-hyperarc consistency (Definition~\ref{def:kconsist}). Despite DRLs form an intensively studied algebraic variety \cite{WD39,DBLP:journals/ijac/BlountT03,JM2006}, they have never been proposed as an evaluation framework for soft constraints, hence we briefly discuss their logical and algebraic motivation.

As already mentioned, soft CSPs are a generalization of crisp CSPs. Conversely, crisp CSPs can be regarded as a particular soft CSPs, evaluated over the algebra $(\{0,1\},\leq,\odot,\bot,\top)$ described above. Since the previous algebra is a reduct of the familiar Boolean algebra $\mathbf{2}$ (taking $\odot$ as $\wedge$), 
and since $\mathbf{2}$ and the meet operation in $\mathbf{2}$ form the algebraic counterparts 
of Boolean logic and Boolean conjunction respectively, it is natural to 
intend the combination operator $\odot$ in a valuation structure 
as a generalization of the meet operation in $\mathbf{2}$ and to investigate 
the algebraic counterparts of logics that generalize Boolean conjuction 
as candidate as valuation structures for soft CSPs. Intriguingly, a central approach in the area of \emph{mathematical fuzzy logic}, popularized by H{\'a}jek \cite{H98}, relies on the idea of generalizing Boolean logic starting from a generalization of Boolean conjunction by means of a class of functions called \emph{(continuous) triangular norms} \cite{KMP00}. The idea is the following. A triangular norm $*$ is an  associative, commutative,  continuous binary function over the real interval $[0,1]$; 
moreover, $*$ is monotone over the (total, dense and complete) order of reals in $[0,1]$, has $1$ as identity and $0$ as annihilator. Given a (continuous) triangular norm $*$, there exists a unique binary function $\rightarrow_*$ on $[0,1]$ satisfying the \emph{residuation} equivalence, 
\begin{equation*}
x * z \leq y \mbox{ if and only if } z \leq x \rightarrow_* y \mbox{,}
\end{equation*}
namely $x \rightarrow_* y = \max \{ z ~|~ x * z \leq y \}$. This function is called \emph{residuum}, 
and is a generalization of the Boolean implication. Thus, on the basis of any triangular norm $*$, 
a corresponding propositional fuzzy logic, $$\mathcal{L}_*=([0,1],\wedge,\vee,\odot,\rightarrow,\neg,\bot,\top)\mbox{,}$$ 
is obtained by interpreting propositional variables over $[0,1]$, 
$\bot$ over $0$, $\odot$ over $*$, $\rightarrow$ over $\rightarrow_*$, 
and eventually by defining $\neg x =x \rightarrow \bot$, $\top = \neg \bot = 1$, 
$x \wedge y = x \odot (x \rightarrow y) = \min(x,y)$, and $x \vee y = ((x \rightarrow y)\rightarrow y) \wedge ((y \rightarrow x)\rightarrow x)= \max(x,y)$. It is immediate to realize that the Boolean logic 
can be recovered from $\mathcal{L}_*$ by restricting the domain and the connectives to $\{ 0,1\}$. 
As much as Boolean algebras form the equivalent algebraic semantics 
of Boolean logic, in the sense of Blok and Pigozzi \cite{BP89}, 
the variety of \emph{$BL$-algebras} (defined in Section~\ref{sect:resLatt}) 
forms the algebraic semantics of the logic of all continuous 
triangular norms and their residua, called \emph{H{\'a}jek's basic logic} \cite{H98,CEGT99}. 

Therefore, $BL$-algebras can be regarded as first candidates as an evaluation framework for soft CSPs. 
However, as far as $BL$-algebras are regarded as hosts for implementing $k$-hyperarc consistency enforcing algorithms, the \emph{prelinearity} equation, 
\begin{equation*}
(x \rightarrow y) \vee (y \rightarrow x) = \top \mbox{,}
\end{equation*}
turns out to be redundant. Since prelinearity is exactly the property that specializes $BL$-algebras 
inside the class of DRLs \cite{JM2006}, we are led to DRLs as a defensible level of generality 
for an evaluation framework. On the logical side, 
the DRLs variety forms the algebraic semantics of an intersecting common 
fragment of basic logic and intuitionistic logic, called \emph{generalized basic logic} \cite{BM07}. 
We insist that, in general, DRLs are not totally ordered nor, with the exception of \emph{G{\"odel} algebras}, idempotent. 

In light of the above, we adopted DRLs as valuation structures for soft constraints, 
and we obtained two results. The first is that preeminent valuation structures for soft constraints turn out to be members of subvarieties of DRLs, namely commutative idempotent semirings are \emph{Heyting algebras} (Proposition~\ref{prop:cis}) and fair valuation structures are totally ordered \emph{$BL$-algebras} (Proposition~\ref{prop:fvs}). 
As a second result, we describe a  polynomial-time algorithm that enforces a natural local consistency property, called \emph{$k$-hyperarc consistency} (Definition~\ref{def:kconsist}), on soft CSPs evaluated over DRLs (Theorem~\ref{theor:main}). This property guarantees that any \emph{consistent} assignment to a variable $i$ extends to an assignment to any other $\leq k-1$ variables constrained by $i$, without producing additional costs. On the one hand we remark that, in contrast with idempotent cases, the closure of a soft CSP under this local consistency property is \emph{not unique}. The notion of \emph{optimal} closure, and the complexity of finding such closures (which is a key property to embed enforcing algorithms into a branch and bound search, \cite{DBLP:conf/ijcai/CooperGS07}), deserve further investigation. On the other hand, we insist that our algorithm works \emph{uniformly} over every DRL, including the aforementioned structures as special cases. For this reason, we expect the DRLs framework to allow for a relatively easy migration of other local consistency techniques that currently work in the crisp and soft settings.

We conclude the introduction with a suggestion for an applicative development of this work. Once valuation structures are established to be algebras in subvarieties of DRLs, universal algebraic considerations guarantee that taking \emph{free} algebras as concrete representatives, for instance in applications, is a suitable choice, since any equation satisfied by the free algebra is satisfied by every algebra in the variety \cite{MKMNT87}. As regards to certain locally finite subvarieties of DRLs, namely G{\"odel} algebras and Komori $MV$-algebras, combinatorial representations of free algebras are available \cite{MR2250542,CdOM99}, and perhaps even more interestingly, combinatorial constructions of free algebras over given finite distributive lattices are known \cite{AGM08,M08}. The latter constructions give the opportunity to fix a suitable lattice structure, depending on the actual soft CSP of interest, and then to construct the most general valuation structure on top of this ordered structure. 

\subsubsection{Outline} The paper is organized as follows. 
In Section~\ref{sect:softCSPs}, we define soft CSPs and valuation structures. 
In Section~\ref{sect:resLatt}, we define divisible residuated lattices, 
and we list a number of properties qualifying DRLs as suitable and natural 
valuation structures for soft constraints. Then, we describe the relation between 
evaluation frameworks such as commutative idempotent semirings and fair valuation structures, 
and DRLs. 
In Section~\ref{sect:hyperArcConsist} we present the main technical contribution of this paper, 
that is a uniform polynomial-time algorithm for $k$-hyperarc consistency enforcing on soft CSPs 
evaluated over DRLs. 

For background notions on partial orders and universal algebra, 
we refer the reader to \cite{DP02} and \cite{MKMNT87} respectively. 

\subsubsection{Acknowledgments} The author thanks Stefano Bistarelli and Vincenzo Marra for fruitful discussions on the subject of this paper.

\section{Soft Constraint Satisfaction Problems} \label{sect:softCSPs}

In this section, we define formally the notions of soft CSPs, valuation structure, 
and optimal solution to a soft CSP. 

A \emph{(soft) constraint satisfaction problem} (in short, \emph{CSP}) 
is a tuple 
$$\mathbf{P}=(X,D,P,\mathbf{A})\mbox{,}$$
specified as follows. 

$X = \{ 1,\dots,n \}=[n]$ is a set of \emph{variables}, 
and $D = \{ D_i \}_{i \in [n]}$ is a set of finite \emph{domains} over which variables are 
assigned, variable $i$ being assigned over domain $D_i$. Let $Y \subseteq X$. 
We let 
$$l(Y)=\prod_{i \in Y}D_i$$ 
denote 
all the assignments of variables in $Y$ onto the corresponding domains (\emph{tuples}). 
If $Y=\emptyset$, then $l(Y)$ contains only the empty tuple. 
For any $Z \subseteq Y$, we denote by $t|_{Z}$ the \emph{projection} of $t$ onto the variables in $Z$. 
For every $i \in Y$, $a \in D_i$ and $t \in l(Y \setminus \{ i\})$, 
we let $t \cdot a$ denote the tuple $t'$ in $l(Y)$ such that 
$t'|_{\{ i\}}=a$ and $t'|_{Y \setminus \{ i\}}=t$ 
(if $Y=\{ i\}$, then $t \cdot a=a$). 

$\mathbf{A}$ is an algebra with domain $A$ and signature including 
a binary relation $\leq$, a binary operation $\odot$ and constants $\top$, $\bot$, such that 
the reduct  $(A,\leq,\top,\bot)$ is a bounded poset (that is, $\leq$ is a partial order with greatest element $\top$ and least element $\bot$), and the reduct $(A,\odot,\top)$ is a commutative monoid (that is, $\odot$ is commutative and associative and has identity $\top$) where $\odot$ is \emph{monotone} over $\leq$, 
that is $x \leq y$ implies $x \odot z \leq y \odot z$. $\mathbf{A}$ is called 
the \emph{valuation structure} of $\mathbf{P}$, 
and $\odot$ is called the \emph{combination} operator over $\mathbf{A}$.

$P$ is a finite multiset 
\footnote{Multisets are necessary to support nonidempotent combinations of constraints.} 
of \emph{constraints}. Each constraint $C_Y \in P$ is defined over a subset $Y \subseteq X$ as a map 
$$C_Y : \prod_{i \in Y}D_i \rightarrow A\mbox{.}$$
A constraint $C_Y$ has \emph{scope} $Y$ and \emph{arity} $|Y|$. 

Let $(C_{Y_1},\dots,C_{Y_m})$ be an $m$-tuple of constraints in $P$, 
and let $f$ be an $m$-ary operation on $A$. Then, $f(C_{Y_1},\dots,C_{Y_m})$ 
is the constraint with scope $Y_1 \cup \dots \cup Y_m$ defined by putting, 
for every $t \in l(Y_1 \cup \dots \cup Y_m)$: 
$$f(C_{Y_1},\dots,C_{Y_m})(t) = f(C_{Y_1}(t|_{Y_1}),\dots,C_{Y_m}(t|_{Y_m}))\mbox{.}$$

The set $S(\mathbf{P})$ of \emph{(optimal) solutions} to $\mathbf{P}$ is equal 
to the set of $t \in l(X)$ such that $\bigodot_{C_Y \in P}C_Y(t|_{Y})$ is \emph{maximal} in the poset:
$$\left\{ \left. \bigodot_{C_Y \in P}C_Y(t|_{Y}) ~\right|~  t \in l(X) \right\} \subseteq A\mbox{,}$$
where an element $x$ is maximal in a poset if there is no element $y > x$ in the poset 
(notice that maximal elements in a poset form an antichain). 
If $S(\mathbf{P})=\{ \bot \}$, we say that $\mathbf{P}$ is \emph{inconsistent}. 

Let $\mathbf{P}=(X,D,P,\mathbf{A})$ and $\mathbf{P}'=(X,D,P',\mathbf{A})$ be CSPs. 
We say that $\mathbf{P}$ and $\mathbf{P}'$ are \emph{equivalent} (in short, $\mathbf{P} \equiv \mathbf{P}'$) 
if and only if for every $t \in l(X)$, 
$$\bigodot_{C_Y \in P}C_Y(t|_{Y}) = \bigodot_{C_Y \in P'}C_Y(t|_{Y}) \mbox{.}$$
In particular, if $\mathbf{P} \equiv \mathbf{P}'$, 
then $S(\mathbf{P})=S(\mathbf{P}')$. 

In the sequel we shall assume the following, without loss of generality: 
$P$ contains at most one constraint with scope $Y \neq \emptyset$ for every $Y \subseteq X$ 
(otherwise, we replace any pair of constraints $C'_Y, C''_Y$ by the constraint $C_Y$ 
defined by $C_Y(t|_Y)=C'_Y(t|_Y) \odot C''_Y(t|_Y)$ for every $t \in l(Y)$); 
$P$ contains all the constraints $C_{\{i\}}$ for $i=1,\dots,n$ (otherwise, 
we add the constraint $C_{\{i\}}$ stipulating that $C_{\{i\}}(a) = \top$ for every $a \in D_i$); 
$C_{\{i\}}(a)>\bot$ for every $a \in D_i$ (otherwise, we remove $a$ from $D_i$, 
declaring the problem inconsistent if $D_i$ becomes empty). 
Moreover, we shall assume that constraints are implemented as tables, 
such that entries can be both retrieved and modified, and that algebraic operations 
over the valuation structure are polynomial-time computable in the size of their inputs. 

\section{Divisible Residuated Lattices} \label{sect:resLatt}

In this section, we introduce the variety of DRLs and some of its subvarieties, 
which are interesting with respect to soft CSPs. We give the logical interpretation of each mentioned algebraic variety, and we formalize the relation between DRLs and, 
commutative idempotent semirings and fair valuation structures. 

\begin{definition}[Divisible Residuated Lattice, DRL] \label{def:drl}
A \emph{divisible residuated lattice} (or \emph{$GBL$-algebra}) 
\footnote{To our aims, we can restrict to commutative and bounded residuated lattices. 
We refer the reader to \cite{JM2006} for a general definition.} 
is an algebra $(A,\vee,\wedge,\odot,\rightarrow,\top,\bot)$ such that:
\begin{enumerate}
\item[$(i)$] $(A,\odot,\top)$ is a commutative monoid; 
\item[$(ii)$] $(A,\vee,\wedge,\top,\bot)$ is a bounded lattice (we write $x \leq y$ 
if and only if $x \wedge y=x$); 
\item[$(iii)$] \emph{residuation} holds, that is: 
\begin{equation}\label{eq:resid}
x \odot z \leq y \mbox{ if and only if } z \leq x \rightarrow y\mbox{;}
\end{equation}
\item[$(iii)$] \emph{divisibility} holds, that is: 
\begin{equation}\label{eq:divis}
%x \leq y \mbox{ implies } x \odot (x \rightarrow y) = y \mbox{.}
x \wedge y = x \odot (x \rightarrow y) \mbox{.}
\end{equation}
\end{enumerate}
A DRL is called a \emph{chain} if its lattice reduct is totally ordered.
\end{definition}

We remark that residuation can be readily rephrased in equational terms, so that DRLs form a variety. 
Notice that chains are not closed under direct products, so that they do not form varieties. 
As a fact, the lattice reduct of a DRL is \emph{distributive}, that is, 
$x \wedge (y \vee z)=(x \wedge y) \vee (x \wedge z)$. 

The monoidal operation of a DRL matches the minimal requirements imposed over the combination operator of a valuation structure in Section~\ref{sect:softCSPs}, as summarized by the following fact \cite{S00}.

\begin{fact}[DRLs Basic Properties] \label{prop:basicOps}
Let $\mathbf{A}$ be a DRL. For every $x,y,z \in A$:
\begin{enumerate}
\item[$(i)$] $x \odot (y \odot z) = (x \odot y) \odot z$.
\item[$(ii)$] $x \odot y = y \odot x$.
\item[$(iii)$] $x \odot \top = x$.
\item[$(iv)$] $x \odot \bot = \bot$.
\item[$(v)$] $x \leq y$ implies $x \odot z \leq y \odot z$ (in particular, $x \odot x \leq x$).
\end{enumerate}
\end{fact}

Some additional properties of DRLs will be useful later \cite{S00}.

\begin{fact}[DRLs Additional Properties] \label{prop:addProp}
Let $\mathbf{A}$ be a DRL. For every $x,y,z \in A$:
\begin{enumerate}
\item[$(i)$] $x \leq y$ if and only if $x \rightarrow y = \top$. 
\item[$(ii)$] $y \leq x$ implies $x \odot (x \rightarrow y) = y$. 
\item[$(iii)$] $y \leq z$ implies $(x \odot z) \odot (z \rightarrow y) = x \odot y$.
\item[$(iv)$] $x \odot (y \vee z) = (x \odot y) \vee (x \odot z)$.
\end{enumerate}
\end{fact}

In the rest of this section, we show that commutative idempotent semirings and fair valuation structures 
are special cases of DRLs, that is, they are algebras in certain subvarieties of DRLs. We first 
introduce the relevant subvarieties of DRLs, and we recall their logical interpretation (see Figure~\ref{fig:drlssubv}). 

\begin{definition}[DRLs Subvarieties] \label{def:subvar}
A \emph{$BL$-algebra} is a \emph{prelinear} $GBL$-algebra, that is, 
\begin{equation} \label{eq:prelin}
(x \rightarrow y) \vee (y \rightarrow x) = \top \mbox{.}
\end{equation}
A \emph{Heyting algebra} is an \emph{idempotent} $GBL$-algebra, that is, 
\begin{equation} \label{eq:idem}
x \odot x = x \mbox{.}
\end{equation}
An \emph{$MV$-algebra} is an \emph{involutive} $BL$-algebra, that is, 
\begin{equation} \label{eq:inv}
\neg \neg x = x \mbox{,}
\end{equation}
where $\neg x = x \rightarrow \bot$. A \emph{G{\"o}del algebra} is an \emph{idempotent} $BL$-algebra. 
A \emph{Boolean algebra} is (equivalently) an involutive Heyting algebra, 
or an idempotent $MV$-algebra, or an involutive G{\"odel} algebra.
\end{definition}

We already mentioned in the introduction that the variety of $BL$-algebras form the equivalent algebraic semantics 
of H{\'a}jek's basic logic. We recall here that the variety of Heyting algebras, $MV$-algebras, G{\"o}del algebras, and Boolean  algebras respectively, form the equivalent algebraic semantics of \emph{intuitionistic logic}, \emph{{\L}ukasiewicz logic}, \emph{G{\"o}del logic}, and \emph{classical logic} \cite{R1974,H98,CdOM99}. As a general fact, if an algebraic variety $V$ forms the algebraic semantic of a propositional logic $\mathcal{L}$, then algebraic properties have a natural logical  counterpart and viceversa. Indeed, the free ($n$-generated) algebra in the variety $V$ is isomorphic to the Lindenbaum-Tarski algebra (of the $n$-variate fragment) of the logic $\mathcal{L}$.

We describe now the relation between DRLs and two very general and intensively investigated evaluation frameworks for soft constraints, namely commutative idempotent semirings and fair valuation structures. As already mentioned, it turns out that the latter can be attained from the former by explicitating structure (defining operations), and the former can be retrived from the latter by forgetting structure (taking reducts). We exploit the following fact. 

\begin{fact} \cite{P79} \label{prop:resid}
Let $(A,\vee,\wedge,\top,\bot)$ be a complete bounded lattice, 
and let $\odot$ be a commutative 
\footnote{Monotonicity of $\odot$ on both arguments is sufficient.} 
monotone operation over $A$ such that $\odot$ distributes over $\vee$. Then, there 
exists a unique operation $x \rightarrow y$ satisfying the residuation equation (\ref{eq:resid}), 
namely for every $x,y,z \in A$, 
$$x \rightarrow y = \bigvee \{ z ~|~ x \odot z \leq y \}\mbox{.}$$
\end{fact}

We consider first commutative idempotent semirings. The restriction to the idempotent case is motivated in this context, since basically local consistency techniques succeed only over idempotent semirings \cite{BMRSVF99}.

\begin{definition}[Commutative Idempotent Semiring, CIS] \label{def:cis}
A \emph{commutative idempotent semiring} 
is an algebra $(A,\vee,\odot,\top,\bot)$ such that:
\begin{enumerate}
\item[$(i)$] $\vee$ is commutative, associative, idempotent, 
$x \vee \bot=x$ and $x \vee \top=\top$;
\item[$(ii)$] $\odot$ is commutative, associative, idempotent, 
$x \odot \top=x$ and $x \odot \bot=\bot$;
\item[$(iii)$] $\odot$ distributes over $\vee$, that is $x \odot (y \vee z) = (x \odot y) \vee (x \odot z)$.
\end{enumerate}
\end{definition}

We exploit the following fact on CISs.

\begin{fact} \label{fact:cis} \cite{BMR97}
Let $\mathbf{A}=(A,\vee,\odot,\top,\bot)$ be a CIS. 
Then, $(A,\vee,\wedge,\top,\bot)$, where $x \wedge y = x \odot y$, 
is a complete bounded lattice.
\end{fact}

\begin{proposition} \label{prop:cis}
Let $\mathbf{A}=(A,\vee,\odot,\top,\bot)$ be a CIS. 
Then, the expansion $\mathbf{A}'=(A,\vee,\wedge,\odot,\rightarrow,\top,\bot)$ of $\mathbf{A}$, 
where $x \wedge y = x \odot y$ and 
$$x \rightarrow y = \bigvee \{ z ~|~ x \odot z \leq y \}\mbox{,}$$ 
is a Heyting algebra. 
\end{proposition}
\begin{proof}
By Definition~\ref{def:cis}$(ii)$, $(A,\odot,\top)$ is a commutative monoid. 
Moreover, By Fact~\ref{fact:cis}, $(A,\vee,\wedge,\top,\bot)$ is a bounded lattice. 
We have to show that residuation and divisibility 
(equations (\ref{eq:resid}) and (\ref{eq:divis}) respectively) 
hold in $\mathbf{A}'$. 

For residuation, since $(A,\vee,\odot,\top,\bot)$ is complete by Fact~\ref{fact:cis}, 
and $\odot$ distributes over $\vee$ by Definition~\ref{def:cis}$(iii)$, 
we have by Fact~\ref{prop:resid} that the operation $\rightarrow$ 
defined above is in fact the unique operation over $A$ satisfying (\ref{eq:resid}). 
For divisibility, we have to show that $x \wedge y=x \odot (x \rightarrow y)$, 
or equivalently that $x \leq y$ implies $x \odot (x \rightarrow y)=x$. 
First recall that 
$x \odot y = x \wedge y$ in $\mathbf{A}'$. Clearly, 
$x \wedge (x \rightarrow y) \leq x$. Moreover, 
$x \leq y$ if and only if $x \odot x\leq y$ by idempotency, 
if and only if $x \leq x \rightarrow y$ by residuation, 
if and only if $x \odot x \leq x \rightarrow y$ by idempotency, 
if and only if $x \leq x \odot (x \rightarrow y)$ by residuation, 
if and only if $x \leq x \wedge (x \rightarrow y)$. We conclude that $\mathbf{A}'$ is a Heyting algebra, 
since $x \odot x=x$ holds in $\mathbf{A}'$ by hypothesis over $\mathbf{A}$. QED
\end{proof}

\begin{proposition}
Let $\mathbf{A}=(A,\vee,\wedge,\odot,\rightarrow,\top,\bot)$ 
be a Boolean algebra. Then, the reduct $\mathbf{A}'=(A,\vee,\odot,\top,\bot)$ of $\mathbf{A}$ is a CIS.
\end{proposition}
\begin{proof}
Notice that $x \odot x = x = x \wedge x$ holds in $\mathbf{A}$. Thus, 
since properties $(i)$-$(iii)$ of Definition~\ref{def:cis} hold in the lattice reduct of $\mathbf{A}$, 
$\mathbf{A}'$ is a CIS. QED
\end{proof}

Next we consider fair valuation structures. We remark that our definition \emph{dualizes} the original of \cite{CS04}, where the structure, $(A,\leq,\oplus,\ominus,\top,\bot)$, is specified by requiring that $(A,\leq,\top,\bot)$ is a bounded chain, $\oplus$ is commutative, associative, monotone, with identity $\bot$ and annihilator $\top$, 
and eventually that, if $x \leq y$, then $y \ominus x = \bigvee\{ z ~|~ x \oplus z = y \}$. 
The proposed dualization is defensible in logical terms, 
since the operation of combining soft constraints is intended as a \emph{conjunction}, and the monoidal operation of a DRL is in fact a generalization of Boolean conjunction. \footnote{In \cite{CS04}, the authors explicitly relate their combination operator with \emph{triangular conorms}, which are in fact dual to triangular norms discussed in the introduction. These operations are customarily intended as generalizations of Boolean \emph{disjunction}.} As a technicality, we also extended the definition of $x \rightarrow y$ over all the domain of structure, and we present the order relation $\leq$ in terms of the lattice operations $\vee$ and $\wedge$ ($x \leq y$ if and only if $x \vee y=y$ if and only if $x \wedge y=x$). 

\begin{definition}[Dual Fair Valuation Structure, FVS] \label{def:fvs}
A \emph{(dual) fair valuation structure} is an algebra $(A,\leq,\odot,\rightarrow,\top,\bot)$ such that:
\begin{enumerate}
\item[$(i)$] $(A,\vee,\wedge,\top,\bot)$ is a bounded complete chain;
\item[$(ii)$] $\odot$ is commutative, associative, monotone, $x \odot \top=x$ and $x \odot \bot=\bot$;
\item[$(iii)$] $x \rightarrow y = \bigvee \{ z ~|~ x \odot z \leq y \}$.
\end{enumerate}
\end{definition}

\begin{proposition} \label{prop:fvs}
Let $\mathbf{A}=(A,\vee,\wedge,\odot,\rightarrow,\top,\bot)$ be a FVS. 
Then, $\mathbf{A}$ is a $BL$-chain.
\end{proposition}
\begin{proof}
By Definition~\ref{def:fvs}, $(A,\vee,\wedge,\top,\bot)$ 
is a bounded chain and $(A,\odot,\top)$ is a commutative monoid. 
We have to show that residuation, divisibility and prelinearity 
(equations (\ref{eq:resid}), (\ref{eq:divis}) and (\ref{eq:prelin}) respectively) 
hold in $\mathbf{A}$. 

For residuation, 
suppose that $x \odot z \leq y$. Then, 
$z \leq \bigvee \{ z ~|~ x \odot z \leq y \} = x \rightarrow y$. 
Conversely, suppose that $z \leq x \rightarrow y$. Then, 
$x \odot z \leq x \odot (x \rightarrow y)$ by monotonicity, 
and $x \odot (x \rightarrow y) \leq y$ by definition of $\rightarrow$ in $\mathbf{A}$. 
For divisibility, we have to show that $x \wedge y=x \odot (x \rightarrow y)$, 
or equivalently that $x \leq y$ implies $x \odot (x \rightarrow y)=x$. 
Suppose $x \leq y$. By definition of $\rightarrow$ in $\mathbf{A}$, 
$x \leq y$ implies $x \rightarrow y = \top$. But then, 
since $x \odot \top = x$ holds in $\mathbf{A}$, 
we conclude that $x \odot (x \rightarrow y)=x$. 
For prelinearity, let $x,y \in A$, 
so that either $\bot \leq x \leq y \leq \top$ or $\bot \leq y \leq x\leq \top$, 
since $(A,\vee,\wedge,\top,\bot)$ is a bounded chain. As before, 
in the former case $x \rightarrow y = \top$ and in the latter case $y \rightarrow x = \top$, 
therefore prelinearity holds in $\mathbf{A}$, since $x \vee \top = \top$. QED
\end{proof}

\begin{proposition}
Let $\mathbf{A}=(A,\vee,\wedge,\odot,\rightarrow,\top,\bot)$ 
be a complete $GBL$-chain. Then, $\mathbf{A}$ is a FVS.
\end{proposition}
\begin{proof}
First note that $(A,\vee,\wedge,\top,\bot)$ is a complete bounded chain by hypothesis. Also, 
$\odot$ is commutative, associative and has identity $\top$ by Definition~\ref{def:drl}, 
and moreover $\odot$ has annihilator $\bot$ and is monotone by Fact~\ref{prop:basicOps}$(iv)$-$(v)$. 
Since $\odot$ distributes over $\vee$ by Fact~\ref{prop:addProp}$(iv)$, 
we have by Fact~\ref{prop:resid} that $x \rightarrow y = \bigvee \{ z ~|~ x \odot z \leq y \}$ in $\mathbf{A}$, 
since this is the unique operation $\rightarrow$ on $A$ satisfying the residuation equation (\ref{eq:resid}). 
QED
\end{proof}

We conclude this section observing that the soft CSP evaluation framework known as \emph{fuzzy CSP} \cite{BMR97}, 
which has the form $([0,1],\vee,\wedge,\top,\bot)$, can be extended to the G{\"o}del chain  $([0,1],\vee,\wedge,\odot,\rightarrow,\top,\bot)$ putting $x \odot y = x \wedge y$ 
and $x \rightarrow y$ equal to $y$ if $y>x$ and to $\top$ otherwise. This chain turns out to be the \emph{generic} algebra in the variety of G{\"o}del algebras.

\section{Enforcing $k$-Hyperarc Consistency on Soft CSPs \\over Divisible Residuated Lattices} \label{sect:hyperArcConsist}

In this section, we define a property of local consistency, called $k$-hyperarc consistency, 
and we describe a polynomial-time algorithm that enforces $k$-hyperarc consistency on soft CSPs 
evaluated over DRLs. In the rest of this section, it is intended that valuation structures are DRLs. 

\begin{definition}[$k$-Hyperarc Consistency] \label{def:kconsist}
Let $\mathbf{P}=(X,D,P,\mathbf{A})$ be a CSP. 
Let $Y \subseteq X$ such that $2 \leq |Y| \leq k$ and $C_Y \in P$. 
We say that $Y$ is \emph{$k$-hyperarc consistent} if 
for each $i \in Y$ and each $a \in D_i$ such that $C_{\{i\}}(a) > \bot$, 
there exists $t \in l(Y\setminus \{ i \})$ such that,
\begin{equation} \label{eq:kHyperConsist} 
C_{\{i\}}(a)=C_{\{i\}}(a) \odot C_Y(t \cdot a)\mbox{.}
\end{equation}
We say that $\mathbf{P}$ is 
\emph{$k$-hyperarc consistent} if every $Y \subseteq X$ such that 
$2 \leq |Y| \leq k$ and $C_Y \in P$ is $k$-hyperarc consistent. 
\end{definition} 

Notice that equation (\ref{eq:kHyperConsist}) holds if $C_Y(t \cdot a)=\top$. 
Thus, $Y$ is $k$-hyperarc consistency if each assignment $a \in D_i$ of variable $i \in Y$ such that $C_{\{i\}}(a) > \bot$, extends to an assignment $t \in l(Y\setminus \{i\})$ of variables $Y\setminus \{i\}$ without producing additional costs. 

The idea beyond enforcing algorithms is to explicitate implicit constraints induced by the problem over certain subsets of variables, thus possibly discovering a \emph{local} unsatisfiability at the level of these variables. As a specialization of this strategy, our algorithm projects costs from constraints of arity greater than one to constraints of arity one, thus it possibly reveals the unsatisfiability of the subproblem induced over a single variable (or else, it possibly shrinks the domain of that variable). Such a local unsatisfiability implies the unsatisfiability of the whole problem, as the following proposition shows. 

\begin{proposition} \label{prop:localinconsist}
Let $\mathbf{P}=(X,D,P,\mathbf{A})$ be a CSP and let $i \in [n]$ 
be such that $C_{\{ i\}} \in P$ and $C_{\{ i\}}(a)=\bot$ 
for every $a \in D_i$. Then, $\mathbf{P}$ is inconsistent.
\end{proposition}
\begin{proof} 
First recall that for every $x \in A$ it holds that $x \odot \bot=\bot$. 
But then, $C_{\{ i\}}(t|_{\{i\}}) = \bot$ for every $t \in l(X)$, 
so that $\bigodot_{C_Y \in P}C_Y(t|_{Y}) = \bot$. Therefore, 
$S(\mathbf{P})=\{ \bot \}$ and $\mathbf{P}$ is inconsistent. QED
\end{proof}

The specifications and the pseudocode of our enforcing algorithm follow. 

\begin{description}
\item[\textbf{Algorithm:}] \textsc{$k$-HyperarcConsistency}
\item[\textbf{Input:}] A CSP $\mathbf{P}=(X,D,P,\mathbf{A})$.
\item[\textbf{Output:}] $\bot$ or a $k$-hyperarc consistent CSP $\mathbf{P}'=(X,D,P',\mathbf{A})$ equivalent to $\mathbf{P}$.
\end{description}

\begin{tabbing}
$k$\textsc{-HyperarcConsistency}$((X,D,P,\mathbf{A}))$\\
1 \quad \= $Q \leftarrow \{ 1,\dots,n\}$ \\
2       \> \textbf{while} $Q \neq \emptyset$ \textbf{do}\\
3       \> \quad \= $i \leftarrow$ \textsc{Pop}$(Q)$\\
4       \>       \> \textbf{foreach} $Y \subseteq X \mbox{ such that } 2 \leq |Y| \leq k \mbox{, } i \in Y \mbox{ and } C_Y \in P$  \textbf{do}\\
5       \>       \> \quad \= domainShrinks $\leftarrow$ \textsc{Project}$(Y,i)$\\
6       \>       \>             \> \textbf{if} $C_{\{i\}}(a)=\bot$ \mbox{ for each } $a \in D_i$ \textbf{then}\\
7       \>       \>             \> \quad \= \textbf{return} $\bot$\\ 
8       \>       \>             \> \textbf{else if} domainShrinks \textbf{then}\\ 
9       \>       \>            \>       \> \textsc{Push}$(Q,i)$\\
10       \>       \>             \> \textbf{endif}\\ 
11       \>       \> \textbf{endforeach}\\
12      \> \textbf{endwhile}\\
13      \> \textbf{return} $(X,D,P',\mathbf{A})$
\end{tabbing}

\begin{tabbing}
\textsc{Project}$(Y,i)$\\
14 \quad \= domainShrinks $\leftarrow$ \textbf{false}\\
15 \quad \= \textbf{foreach} $a \in D_i$ \mbox{such that} $C_{\{i\}}(a)>\bot$ \textbf{do}\\
16       \> \quad \= $x \leftarrow $ \text{a maximal element in} $\{C_Y(t \cdot a)~|~t \in l(Y\setminus \{i\})\}$\\
17       \>       \> $C_{\{i\}}(a) \leftarrow C_{\{i\}}(a) \odot x$\\
18       \>       \> \textbf{if} $C_{\{i\}}(a)=\bot$ \textbf{then}\\
19       \>       \> \quad \= domainShrinks $\leftarrow$ \textbf{true}\\
20       \>       \> \textbf{endif}\\
21       \>       \> \textbf{foreach} $t \in l(Y\setminus \{i\})$ \textbf{do}\\
22       \>       \> \quad \= $C_Y(t \cdot a) \leftarrow (x \rightarrow C_Y(t \cdot a))$\\
23       \>       \> \textbf{endforeach}\\
24      \> \textbf{endforeach}\\
25      \> \textbf{return} domainShrinks
\end{tabbing}

As already discussed in the introduction, enforcing $k$-hyperarc consistency over the $k$-hyperarc inconsistent problem $\mathbf{P}$ may return in output several distinct $k$-hyperarc consistent problems, depending on the choices made on Lines~1, 3, 4 and 16. 

In the rest of this section, we prove that the algorithm runs in polynomial-time (Lemma~\ref{lemma:ptime}) 
and is sound (Lemma~\ref{lemma:sound}). 

\begin{lemma}[Complexity] \label{lemma:ptime}
Let $\mathbf{P}=(X,D,P,\mathbf{A})$ be a CSP, 
where $X=[n]$, $d=\max_{i \in [n]}|D_i|$ and $e=|P|$. Then, 
$k\textsc{-HyperarcConsistency}$ terminates in at most $O(e^2 \cdot d^{k+1})$ time.
\end{lemma}
\begin{proof}
The main loop in Lines~2-12 iterates at most $n(d+1) \leq e(d+1)$ times, since $n \leq e$ without loss of generality and each $i \in [n]$ is added to $Q$ once on Line~1 and at most $d$ times on Line~9 (once for each shrink of domain $D_i$ of size $\leq d$). Each iteration of the main loop involves at most $e$ iterations of the loop nested in Lines~4-11, since there are at most 
$e$ constraints satisfying the condition in Line~4 with respect to any given $i \in [n]$. Each such nested iteration amounts to an invocation of $\textsc{Project}$ and an iteration over domain $D_i$ of size $\leq d$. 
Any invocation of $\textsc{Project}$ amounts to an iteration over domain $D_i$ of size $\leq d$ on Line~15, 
and for each such iteration, two iterations over all the $\leq d^{k-1}$ tuples $t \in l(Y\setminus \{i\})$, 
observing that $1 \leq |Y\setminus \{i\}| \leq k-1$ (Line~16 and Lines~21-23). Summarizing, the algorithm executes at most
$$(e(d+1))e(d + d(2d^{k-1}))$$
many iterations, so that the algorithm terminates in at most $O(e^2 \cdot d^{k+1})$ time. QED
\end{proof}

\begin{lemma}[Soundness] \label{lemma:sound}
Let $\mathbf{P}=(X,D,P,\mathbf{A})$ be a CSP, and consider the output of $k\textsc{-HyperarcConsistency}(\mathbf{P})$:
\begin{enumerate}
\item[$(i)$] if it is $\bot$, then $\mathbf{P}$ is inconsistent; 
\item[$(ii)$] otherwise, it is a $k$-hyperarc consistent CSP equivalent to $\mathbf{P}$.
\end{enumerate}
\end{lemma}
\begin{proof} 
First we show that the subprocedure \textsc{Project} preserves equivalence, in the following sense. 
Let $R'$ be the multiset of constraints before the $j$th invocation of \textsc{Project} in Line~5, 
let $Y$ and $i$ be the parameters of such invocation, 
and let $R''$ be the multiset of constraints computed by the $j$th execution of Lines~14-25. 
We aim to show that for every $t \in l(X)$,
\begin{equation} \label{eq:projectEquivalence}
\bigodot_{C_Y \in R'}C_Y(t|_{Y})=\bigodot_{C_Y \in R''}C_Y(t|_{Y})\mbox{,}
\end{equation}
that is, problems $(X,D,R',\mathbf{A})$ and $(X,D,R'',\mathbf{A})$ are equivalent. 

% Recall that we are assuming for simplicity that $R'$ contains $C_{\{i\}}$ 
% and at most one constraint of the form $C_Y$. Note that \textsc{Project} preserves 
% this property, so that also $P''$ contains $C_{\{i\}}$ and at most one constraint of the form $C_Y$. 
Let $t \in l(X)$ and let $t|_{\{i\}}=a \in D_i$ such that $C_{\{i\}}(a)>\bot$ (Line~15). 
Clearly, $t|_{Y\setminus \{ i\}} \in l(Y\setminus \{i\})$. In Line~16, 
$x$ is settled to a maximal element in the poset 
$$\{C_Y(t|_{Y\setminus \{i\}} \cdot t|_{\{i\}}) ~|~ t|_{Y\setminus \{ i\}} \in l(Y\setminus \{i\})\}\mbox{,}$$
so that by construction $C_{Y}(t|_Y) \leq x$. 
By Line~17, the constraint $C_{\{i\}}(t|_{\{i\}})$ in $R'$ becomes  
$$C_{\{i\}}(t|_{\{i\}}) \odot x$$ 
in $R''$, and by Line~22, at some iteration of the loop of Lines~21-23, 
the constraint $C_{Y}(t|_Y)$ in $R'$ becomes 
$$x \rightarrow C_{Y}(t|_Y)$$ 
in $R''$. Now, we claim that: 
$$C_{\{i\}}(t|_{\{i\}}) \odot C_{Y}(t|_Y) = (C_{\{i\}}(t|_{\{i\}}) \odot x) \odot (x \rightarrow C_{Y}(t|_Y)) \mbox{.}$$
Indeed, in light of Fact~\ref{prop:addProp}$(iii)$ and the aforementioned fact that $C_{Y}(t|_Y) \leq x$,
\begin{align*}
(C_{\{i\}}(t|_{\{i\}}) \odot x) \odot (x \rightarrow C_{Y}(t|_Y)) & = C_{\{i\}}(t|_{\{i\}}) \odot (x \odot (x \rightarrow C_{Y}(t|_Y)))\\
                                    & = C_{\{i\}}(t|_{\{i\}}) \odot (x \wedge C_{Y}(t|_Y))\\
                                    & = C_{\{i\}}(t|_{\{i\}}) \odot C_{Y}(t|_Y)\mbox{.}
\end{align*}
Eventually, \textsc{Project} does not modify constraints $C_Z \in R'$ 
such that $Z \neq \{i\}$ and $Z \neq Y$ , so that,
$$\bigodot_{C_Z \in R',Z \neq \{i\},Z \neq Y}C_Z(t|_{Z})=\bigodot_{C_Z \in R'',Z \neq \{i\},Z \neq Y}C_Z(t|_{Z})\mbox{.}$$ 
Thus, since $z'=z''$ implies $z \odot z'=z \odot z''$ in $\mathbf{A}$ for every $z,z',z'' \in A$ 
by Fact~\ref{prop:basicOps}$(v)$, we conclude that (\ref{eq:projectEquivalence}) holds.

Now suppose that the algorithm outputs $\bot$ in Line~7. 
We claim that the input problem $\mathbf{P}=(X,D,P,\mathbf{A})$ is inconsistent. Indeed, 
let $j$ be such that after the $j$th execution of \textsc{Project}, 
say over parameters $Y$ and $i$, it holds that $C_{\{i\}}(a)=\bot$ for each $a \in D_i$. 
Let $P'$ be the multiset of constraints computed by such $j$th execution. 
Since \textsc{Project} preserves equivalence, $\mathbf{P}'=(X,D,P',\mathbf{A})$ 
is equivalent to $\mathbf{P}$. But, by Proposition~\ref{prop:localinconsist}, 
$\mathbf{P}'$ is inconsistent, so that $\mathbf{P}$ is inconsistent too. 

Next suppose that the algorithm outputs $\mathbf{P}'=(X,D,P',\mathbf{A})$ in Line~13. 
We claim that the output problem is $k$-hyperarc consistent and equivalent to the input problem $\mathbf{P}=(X,D,P,\mathbf{A})$. 
For equivalence, simply note that \textsc{Project} preserves equivalence. For $k$-hyperarc consistency, 
first note that every $i \in [n]$ is such that $C_{\{i\}}(a) > \bot$ for some $a \in D_i$. 
Indeed, this holds in the input problem $\mathbf{P}$ without loss of generality, 
and each execution of \textsc{Project}, which possibly pushes some $C_{\{i\}}(a)$ down to $\bot$ 
in Line~17, is followed by the check of Lines~18-20. 

Now, let $Y \subseteq X$ be such that $2 \leq |Y| \leq k$, $i \in Y$ and $C_Y \in P'$, 
and let $a \in D_i$ be such that $C_{\{i\}}(a)>\bot$. We claim that there exists 
$t \in l(Y \setminus \{ i\})$ such that 
$$C_{\{i\}}(a)=C_{\{i\}}(a) \odot C_Y(t \cdot a)\mbox{.}$$
Note that, by Fact~\ref{prop:basicOps}$(iii)$, equality holds if $C_Y(t \cdot a)=\top$. 
Let $R'$ and $R''$ be respectively 
the multisets of constraints before and after the last execution of 
\textsc{Project} on input $Y$ and $i$. Let $t \in l(Y\setminus \{i\})$ 
be such that $C_Y(t \cdot a)$ is the maximal element in $\{C_Y(t \cdot a)~|~t \in l(Y\setminus \{i\})\}$ 
assigned to $x$ in Line~16. Thus, at some iteration of loop in Lines~21-23, 
we have that the constraint $C_Y(t \cdot a)$ in $R'$ is updated to $x \rightarrow C_Y(t \cdot a)$ in $R''$. But, by Fact~\ref{prop:addProp}$(ii)$, 
$$x \rightarrow C_Y(t \cdot a) = C_Y(t \cdot a) \rightarrow C_Y(t \cdot a)=\top \mbox{,}$$ 
therefore, $C_Y(t \cdot a)=\top$ in $R''$. Noticing that subsequent assignments to 
$C_Y(t \cdot a)$ during the main loop have the form $x \rightarrow \top$, which is equal to $\top$ by Fact~\ref{prop:addProp}$(ii)$, the claim is settled. QED
\end{proof}

\begin{theorem} \label{theor:main}
Let $\mathbf{P}$ be a CSP, and let $\mathbf{P}'=k\textsc{-Hyperarc-Consistency}(\mathbf{P})$. 
Then, $\mathbf{P}'$ is a $k$-hyperarc consistent CSP equivalent to $\mathbf{P}$, 
computed in polynomial time in the size of $\mathbf{P}$.
\end{theorem}

\newcommand{\etalchar}[1]{$^{#1}$}

%%%%%%%%%%%%%%%%%%%%%%%%%%%%%%%%%%%%%%%

\end{document}